\documentclass[12pt,preprint]{aastex}

\begin{document}
\slugcomment{Submitted to ApJ Letter}
\title{The optical counterpart of a Ultraluminous X-Ray Object in M81}

\author{Ji-Feng Liu, Joel N. Bregman and Patrick Seitzer}
\affil{Astronomy Department, University of Michigan, MI 48109}

\begin{abstract}

Ultraluminous X-Ray Objects are off-nucleus point sources with
$L_X=10^{39}$-$10^{41}$ erg/s and are among the most poorly understood 
X-ray sources.  To help understand their nature, we are trying to identify
their optical counterparts by combining images from the Hubble Space Telescope
and the Chandra Observatory. Here we report an optical counterpart for ULX NGC
3031 X-11, which has average X-ray luminosity of $\sim2\times10^{39}$ erg/s and
has varied by a factor of 40\% over the last 20 years. We find a unique optical
counterpart with the magnitude and color of an O8V star and we identify this as
the secondary in a binary system.  The primary is believed to be a black hole
of approximately 18 $M_\odot$, based on analyses of ACIS and ASCA spectra.
This binary system probably is powered by mass transfer from the O8V star onto
the black hole, with the Roche Lobe equal to the stellar radius.  This model
predicts an orbital period of $\sim1.8$ days, which can be tested by future
observations.

\end{abstract} 


\section{INTRODUCTION}

X-ray emission from the nuclei of AGN and quasars, powered by accretion onto
nuclear supermassive black holes, are usually $>10^{41}$ erg/s, while for the
stellar binary systems in the Milky Way, the luminosities are typically
$10^{33}$-$10^{38}$ erg/s.  X-ray objects with luminosities of
$10^{39}$-$10^{41}$ erg/s have been observed in some external galaxies as
off-nucleus point sources (Fabbiano 1989; Marston et al. 1995; Zezas et al.
2002). These objects have several names, the most popular designation being
``Ultraluminous X-ray Objects'', or ULXs.

One suggestion for the nature of these ULXs is that they are binary systems
with $10^3$-$10^4$ $M_{\odot}$ black holes as primaries ( Colbert and Mushotzky
1999). This suggestion is consistent with the X-ray spectra analyses of some
ULXs (Makishima et al. 2000).  However, the formation of such massive black
holes is not predicted by stellar evolution theory and it may be impossible to
form such objects even in dense star clusters.  Alternatively, these sources
may be stellar mass black holes or neutron stars whose emission is beamed,
thereby representing micro-quasars (King et al.  2001). If the emission is
beamed, the intrinsic luminosities become sub-Eddington, and there are known
examples of beamed Galactic X-ray sources. It is also possible that the
luminosities are truly super-Eddington, for example, obtainable from accretion
disks with radiation-driven inhomogeneities (Begelman, 2002).

An essential step in uncovering the nature of the ULXs is to identify them at
other wavelengths, such as optical. Associations of ULXs to emission-line
nebulae based on ROSAT positions have already been done for a number of ULXs
(e.g., Pakull \& Mirioni, 2002).  However, identifications down to single point
sources were impossible given the large positional errors of EINSTEIN and ROSAT
observations, with the smallest error circles had radii of
$\sim5^{\prime\prime}$ (for ROSAT HRI).  With the advent of Chandra X-ray
satellite, this task becomes possible owing to the sub-arcsecond spatial
resolution of Chandra observations. For example, by comparing Chandra ACIS-S
and HST/WFPC2 observations of the ''Antennae'' starburst galaxies, Zezas et al.
(2002) found that 10 out of 14 ULXs are associated with stellar clusters, with
offsets between the X-ray sources and the clusters less than
$2^{\prime\prime}$.  Wu et al.  (2002) found that the ULX in edge-on spiral
galaxy NGC4565 is associated with a stellar cluster on the outskirt of the
bulge.  Goad et al. (2002) found that within the $0\farcs6$ positional error
circle of the ULX in NGC5204, there are three blue objects which they argue are
very young stellar clusters.

As part of our search for optical counterparts of ULXs in nearby galaxies (Liu
et al. 2001), we report the results for a ULX in NGC 3031, NGC 3031 X-11 as
designated in Roberts \& Warwick (2000). This ULX has been observed more than
20 times by EINSTEIN, ROSAT and Chandra X-ray satellites.  Over the 20 year
baseline, its X-ray luminosity varied by more than 40\%, with an average of
$\sim2\times10^{39}$ erg/s.  Matonick \& Fesen (1997) found that this ULX is
coincident with their SNR No. 22 and a discrete radio source in a VLA 20cm map,
and suggested that it may be a SNR, despite its long term variability and high
luminosity.  In $\S$2, we present the observations utilized, the data analysis
procedures, and the results. We discuss the implications of its optical
identification as an O8V star in $\S$3.  For the distance to M81, we use 3.63
Mpc ($\mu=27.8\pm0.2$, Freedman et al.  1994). At such a distance, 1 pixel in
the Planetary Camera chip, i.e., $0\farcs05$, corresponds to a physical scale
of 0.9 parsecs.

\section{OBSERVATIONS AND DATA ANALYSIS}

NGC 3031 X-11 was observed by HST/WFPC2 in June, 2001 for our program. Three
filters, F450W(B), F555W(V), and F814W(I) were used; for each filter, four
500-second exposures were made. We used HSTphot (Dolphin 2000) to perform PSF
fitting photometry. To transform between WFPC2 image positions (X,Y) and
celestial coordinates (RA,DEC), we used IRAF tasks {\tt metric/invmetric},
which correct the geometric distortion introduced by the camera optics. The
final relative positions are accurate to better than $0\farcs005$ for targets
contained on one chip, and $0\farcs1$ for targets on different chips. The
absolute accuracy of the positions obtained from WFPC2 images is typically
$0\farcs5$ R.M.S. in each coordinate (HST data handbook). 

A 50 kilosecond Chandra ACIS observation of NGC 3031 (ObsID 735, observed date:
2000-05-07) was retrieved from the Chandra archive. {\tt CIAO 2.2.1}, {\tt
CALDB 2.12} and {\tt xspec 11.2.0} were used to analyze the ACIS data. Pixel
randomization of event positions was removed to recover the original positions.
{\tt Wavdetect} was run to detect discrete sources on ACIS-S chips. The
accuracy of absolute positions for these sources has a typical R.M.S of
$0\farcs6$ (Aldcroft et al. 2000).

Two bright X-ray sources fall into the WFPC2 field of view, i.e., the ULX on
PC1 and SN 1993J on WF4, with a seperation of $\sim2^\prime$. Using the
relative position of these two sources, we can register the ULX onto the
Planetary Chip very accurately.  We list in Table 1 the ACIS and WFPC2
positions of SN 1993J, the ACIS position for the ULX, and its derived WFPC2
position. The centroiding error for ACIS-S3 positions computed by {\tt
Wavdetect} as quoted in Table 1 is negligible.  The relative positional error
comes mainly from the ACIS plate scale variation\footnote{see
http://cxc.harvard.edu/cal/Hrma/hrma/optaxis/platescale/geom\_public.html},
which produces an error of $0\farcs1$ for a $2^\prime$ seperation, and the
WFPC2 interchip error of $0\farcs1$.  The rotation R.M.S. for both WFPC2 and
ACIS-S chips are $\sim0\fdg01$, and the resulted positional error is about
$0\farcs02$.  The final positional error is $\sim0\farcs15$, i.e., about 3
Planetary Chip pixels.  Within this error circle, there is only one object,
which is indicated by a cross in Figure 1.  

A cluster of typical size 2--3 parsecs, or a supernova remnant, can be readily
resolved in the PC image.  Careful examination shows that this unique
counterpart of NGC 3031 X-11 is not extended, but a point source, with standard
magnitudes $B=23.87\pm0.04$, $V=23.89\pm0.03$, and $I=23.76\pm0.07$ as obtained
with HSTphot.  The counterpart is located on a dusty spiral arm and in a region
of high stellar density, as can be seen in Figure 1.  The Galactic extinction
toward NGC 3031 are $A_B=0.35$, $A_V=0.27$, and $A_I=0.16$, with a HI column
density\footnote{http://heasarc.gsfc.nasa.gov/cgi-bin/Tools/w3nh/w3nh.pl} of
$4.1\times10^{20}$ cm$^{-2}$ (Schlegel et al. 1998). When this extinction is
removed, the counterpart has $M_B=-4.28$, $M_V=-4.18$, $M_I=-4.20$,
$\mbox{(B-V)}_0=-0.10$, and $\mbox{(V-I)}_0=0.02$.  While its absolute
magnitude is consistent with a O9V star (19 $M_\odot$, 7.8 $R_\odot$;
Schmidt-Kaler 1982), its color is much redder, which may indicate intrinsic
extinction by its dusty environments. If we assume an intrinsic
$\mbox{E(B-V)}=0.22$ and a Galactic-like $R_V=3.1$, the counterpart will have
$M_V=-4.9$ and $\mbox{B-V}=-0.32$, matching the color and absolute magnitude of
an O8V star (23 $M_\odot$, 8.5 $R_\odot$), though the I band is dimmer by 0.1
magnitude.  We note that the total color excess $\mbox{E(B-V)}=0.3$ is
consistent with the absorbing column density $2.1\times10^{21}$ cm$^{-2}$ derived
from X-ray spectra according to a Galactic empirical relation
$N_H=5.8\times10^{21}*E(B-V)$ (Bohlin et al. 1979). This star is not a
supergiant, since a supergiant is much redder and much brighter, for example, a
B2I supergiant has $B-V=-0.17$ and $M_V=-6.4$.

\section{DISCUSSION}

The above determination of the mass and radius of the optical counterpart is
based upon comparison with Galactic stars whose properties are known.
Alternatively, one can try to obtain these quantities theoretically by
comparing its observational properties to isochrones (Z=0.020) from the Geneva
stellar models (Lejeune et al. 2001), such as in Figure 2, where we also show
other bright stars in the field. The ages of the bright field stars are from
about $10^6$ to $10^8$ years, with the anticipated implication that star
formation in the spiral arm is not co-eval.  The location of the counterpart in
the color-magnitude diagram is consistent with it being a star younger than
$10^{6.5}$ years.  If it has spent $10^{6.5}$ years on the main sequence, its
absolute magnitude corresponds to that of a star with a mass of 27 $M_\odot$
and a radius of 10.0 $R_\odot$.  A limiting case is that it is a star that has
just reached the main sequence, which makes it hotter and more massive, with a
mass of 41 $M_\odot$ and a radius of 8.7  $R_\odot$.  The stellar metalicity
may differ from solar, which will lead to small changes in the inferred mass
and radius, so we consider two extreme cases of low and high metallicity.  For
a star of metalicity Z=0.004, the age would be less than $10^{6.7}$ years from
its location on the Main Sequence, and its mass ranges from 50 to 28 $M_\odot$
with a radius between 8.3 to 10.3 $R_\odot$ with increasing age.  For a star of
metalicity Z=0.040, it would be younger than $10^{6.4}$ years, its mass lies
between 36 and 26 $M_\odot$ with a radius between 8.8 and 9.8 $R_\odot$.
Taking into account the range of possibilities, the inferred mass is between 26
and 50 $M_\odot$, and the radius is between 8.3 and 10.3 $R_\odot$. This result
can be compared to the mass and radius inferred from its spectral type, i.e.,
23 $M_\odot$ and 8.5 $R_\odot$ for an O8V star. We conclude that the radius is
about $9.3\pm1.0$ $R_\odot$, while the mass is somewhat uncertain, ranging from
23 to 50 $M_\odot$.  The lower end of this mass range is preferred because it
is unlikely that the star has just reached the main sequence since its binary
companion has already evolved into a compact object.  This optical counterpart
is presumably a secondary in a binary system, and is contributing material onto
an accreting compact primary. 

Swartz et al. (2002) analyzed the radial profile of this ULX in their Chandra
ACIS-S3 observation, and found it consistent with a point source; it is
consistent with a point source in our WFPC2 observations.  They also analyzed
the spectrum and found, after deconvolving the effects of pileup, the spectrum
is best described by a model for an absorbed blackbody disk. The best-fit
absorbing column density is $N_{20}=21.7\pm1.0$ in units of $10^{20}$ cm$^{-2}$;
the best-fit innermost disk temperature is $T_{in}=1.03\pm0.11$ Kev, the
corresponding radius is $R_{in}=161\pm16$ km, which correspond to an 18
$M_\odot$ non-rotating black hole. From an analysis of ASCA spectra, Mizuno
 (2000) found the following parameters, $N_{20}=21\pm3$,
$T_{in}=1.48\pm0.08$ Kev, $R_{in}=83\pm8$ km, which he argued correspond to a
Kerr black hole with similar mass.  In Figure 3, we compared the fluxes of the
counterpart in B, V, and I bands with the predictions of the absorbed disk
blackbody model derived from Chandra ACIS and ASCA spectra, and found that the
observed optical fluxes are at least 10-30 times brighter than the disk itself,
indicating that most of the optical flux is from the X-ray inactive secondary
O8V star. Furthermore, the optical disk has a color of $\mbox{B-V}\approx0.1$,
which is significantly flatter (redder) than the color of the counterpart.
These two pieces of evidence lead us to conclude that the optical light is from
a star rather than from an accretion disk.

The nature of the X-ray variability, the X-ray spectral properties, and the
optical colors indicate that NGC 3031 X-11 is not a supernova remnant (Matonick
\& Fesen 1997) but is a high mass X-ray binary (HMXB) system, consisting of an
$M_1=18M_\odot$ black hole and an $M_2=23M_\odot$ O8V companion.  HMXB systems
in our Galaxy with black hole masses above about 10 $M_\odot$ have
$L_X<10^{39}$ erg/s, with very few exceptions.  An example is Cygnus X-1, which
has a BH of 15 $M_\odot$ and a companion of O9.7Iab supergiant, with
$L_X<10^{38}$ erg/s (Tanaka \& Lewin, 1995).  These modest luminosities can be
understood given the low accretion efficiency of stellar wind from the
companion supergiants.  For this ULX system, since the companion is a main
sequence star, its stellar wind is not as strong as supergiants, and it may be
accreting gas via Roche Lobe Overflow. 

This scenario can be tested by future observations. Black hole masses derived
from spectral analyses are best fits to simple models, so changes to the
models, especially for the inner region, may lead to different masses.  In
addition, there is the concern that stellar evolutionary models do not predict
$18 M_\odot$ black holes. For example, models for 60 $M_{\odot}$ main sequence
stars at the point of core collapse (Vanbeveren et al. 1998, and references
therein) predict a mass of $\sim 11 M_\odot$.  These uncertainties in the mass
determination can be avoided by obtaining a direct measure of black hole mass
(function) through measurements of an orbital period and the associated orbital
velocities.  

We can predict the period of the binary based on the mass of the secondary and
the requirement that it is filling its Roche lobe.  If the counterpart has the
radius and mass of an O8V star, i.e., a radius of $\sim8.5R_\odot$ and a mass
of 23 $M_{\odot}$, we infer that the seperation between the primary and
secondary is $a=(0.6q^{2/3}+ln(1+q^{1/3}))q^{-2/3}/0.49R_{cr}\approx21R_\odot$
for their mass ratio $q=M_2/M_1\approx1.3$ (Eggleton 1983). This in turn
implies an orbital period of $P_{orb} \approx1.8$ days according to Kepler's
3rd law.  For the mass range of 23-50 $M_\odot$ and the radius range of
$9.3\pm1.0$ $R_\odot$, the expected orbital period ranges from 1.1 days to 2.1
days.  Although challenging, the measurement of the period would greatly
illuminate the nature of this, and presumably other ULX systems.

King et al. (2001) raised the possibility that ULXs could be HMXB systems during
their short thermal timescale mass transfer via Roche Lobe overflow, and our
identification of ULX NGC 3031 X-11 with an O8V star supports this idea. Such
high mass systems have very short ages, consistent with the general findings
that ULXs tend to occur in star forming regions (e.g., Zezas et al. 2002).
However, ULXs also occur in some elliptical galaxies, e.g., NGC 1399 (Angelini
et al. 2001), and those may be intrinsically different from what we
discuss here.

\acknowledgements

We are grateful for the service of Chandra Data Archive. We would like to thank
Jimmy Irwin, Renato Dupke, Eric Miller, Rick Rothschild, and Steven Murray for
helpful discussions.  We gratefully acknowledge support for this work from NASA
under grants HST-GO-09073.

\begin{deluxetable}{lcccccccccc}
\tablecaption{The optical counterpart for ULX NGC 3031 X-11}
\tabletypesize{\tiny}
\tablehead{
\colhead{} & \multicolumn{4}{c}{ACIS-S3} &\colhead{}& \multicolumn{5}{c}{WFPC2} \\
\cline{2-5}  \cline{7-11} \\
\colhead{Object} & \colhead{RA} & \colhead{DEC} & \colhead{RA\_ERR} & 
\colhead{DEC\_ERR} & \colhead{}&\colhead{Chip} & \colhead{X} & \colhead{Y} &
\colhead{RA} & \colhead{DEC} \\
\colhead{} & \colhead{} & \colhead{} & \colhead{(arcsec)} &
\colhead{(arcsec)} &\colhead{}& \colhead{} & \colhead{(pix)} & \colhead{(pix)} &
\colhead{} & \colhead{}
}

\startdata    
SN 1993J      & 9:55:24.79 & 69:01:13.44 & 0.03 & 0.01 && 4 & 480.72 & 240.89 & 9:55:25.11 & 69:01:14.87 \\ 
NGC 3031 X-11 & 9:55:32.95 & 69:00:33.36 & 0.03 & 0.01 && 1 & 437.57 & 408.32 & 9:55:33.27 & 69:00:34.79 \\

\enddata
\end{deluxetable}

\begin{figure}
\plotone{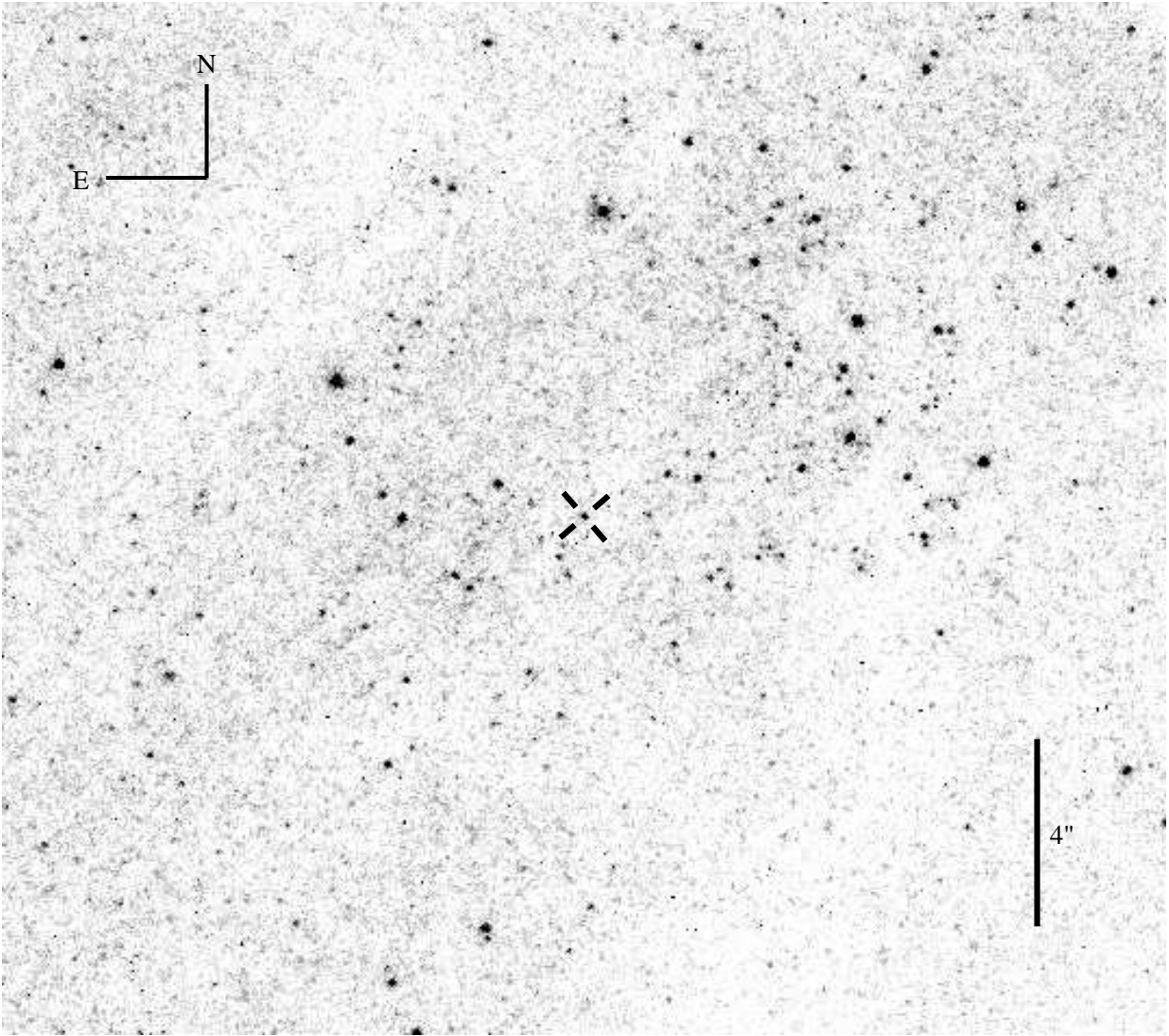}
\caption{The optical counterpart of NGC 3031 X-11 and its environments. The
counterpart, indicated by a cross, is a point source with standard magnitudes
$B=23.87\pm0.04$, $V=23.89\pm0.03$, and $I=23.76\pm0.07$. When the Galactic and
intrinsic extinction is removed, its absolute magnitude and colors correspond
to those of an O8V star.  The size of the image is about
$22^{\prime\prime}\times22^{\prime\prime}$ and the distance between the inner
edges of two diagonal line segments of the cross is twice the diameter of the
positional error circle of the counterpart.  } 

\end{figure}

\begin{figure}
\plotone{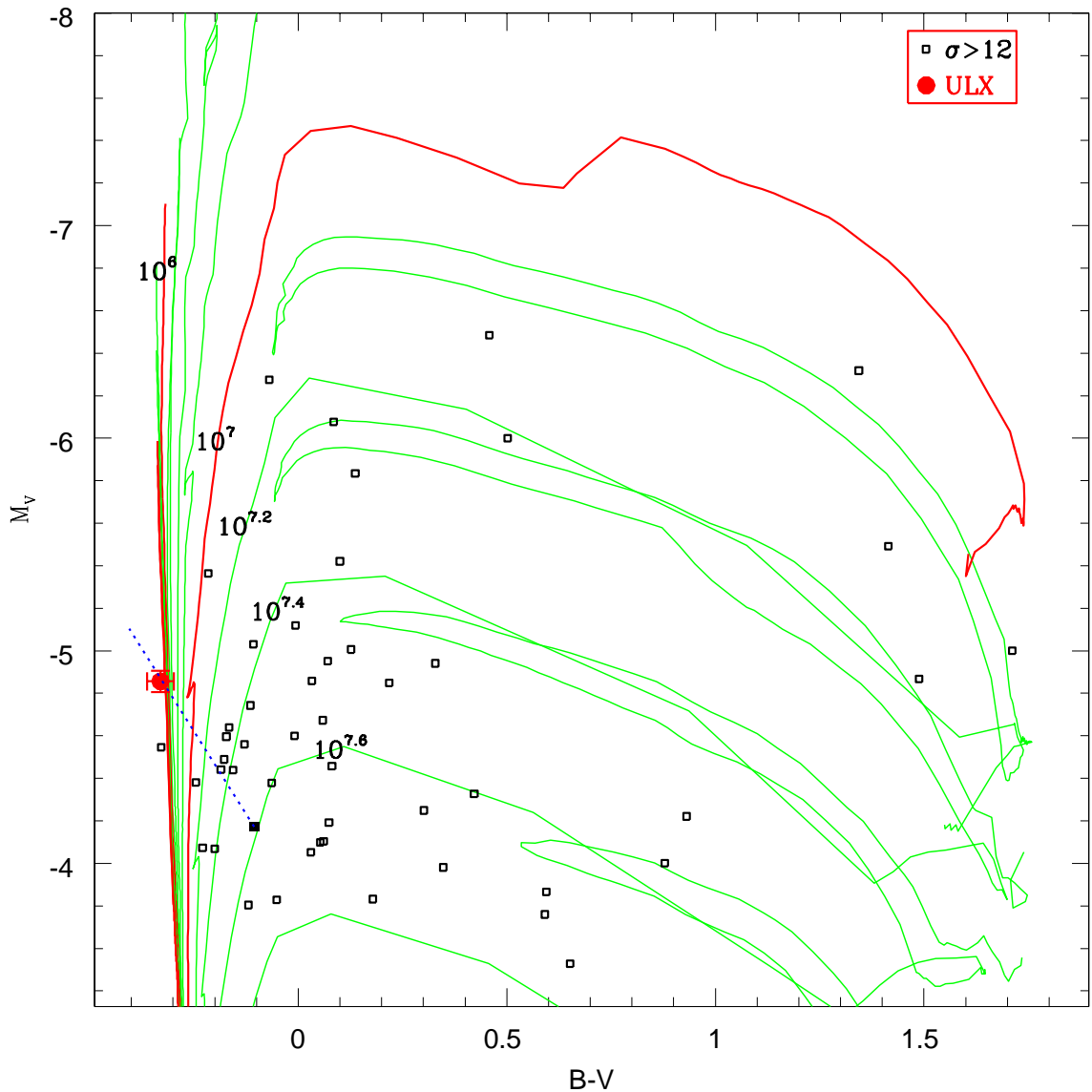}
\caption{Color magnitude diagram for the ULX and nearby bright stars with
signal-to-noise ratio $\sigma>12$. Galactic extinction is removed from all the
stars and the counterpart also is corrected for intrinsic extinction, as
inferred from its X-ray absorption.  The stellar isochrones are taken from
Lejeune et al. (2002) for Z=0.020 and the dashed line indicates the
de-reddening vector for HI absorption assuming $R_V=3.1$.}

\end{figure}

\begin{figure}
\plotone{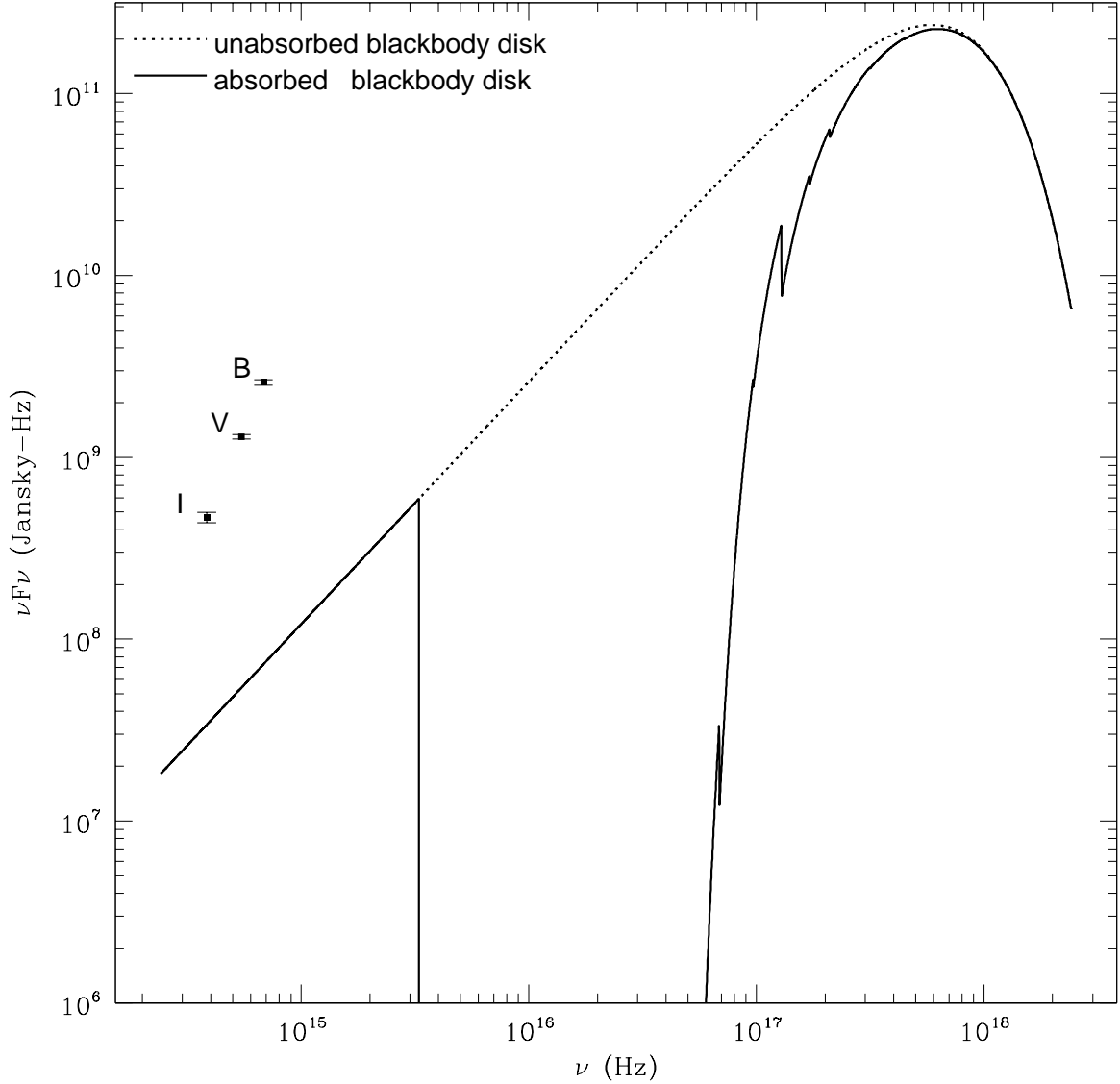}
\caption{Spectral energy distribution of NGC 3031 X-11. The X-ray spectra from
ACIS can be fitted by an absorbed disk blackbody model, which we extend down to
optical to compare with the B, V, and I photometry of the counterpart.  Both
the model and the BVI magnitudes are dereddened.  The BVI fluxes of the
counterpart are 30-10 times larger than the disk itself.}

\end{figure}

\end{document}